\documentclass[10pt]{iopart}
\usepackage{algorithm,algorithmic}
\usepackage{enumerate}
\usepackage{listings}
\usepackage{color}
\usepackage{float}
\usepackage{amssymb}
\usepackage{breqn}
\usepackage{graphicx}

\newcommand{\ket}[1]{\left| #1 \right>} % for Dirac bras
\definecolor{red}{rgb}{1,0.,0}

\newcommand{\revised}[1]{\textcolor{black}{#1}}

\begin{document}

\title[A Computational Workflow for Designing Silicon Donor Qubits]{A Computational Workflow for Designing Silicon Donor Qubits}

\author{Travis S.~Humble$^{1,2,3}$, M.~Nance Ericson$^{1,4}$, \\ Jacek Jakowski$^{1,2,5}$, Jingsong Huang$^{1,2,5}$, \\ Charles Britton$^{1,6}$, Franklin G.~Curtis$^{1,7}$, \\ Eugene F.~Dumitrescu$^{1,2,3}$, Fahd A. Mohiyaddin$^{1,2}$, \\ and Bobby G.~Sumpter$^{1,2,5}$}
\address{$^1$Quantum Computing Institute, Oak Ridge National Laboratory, Oak Ridge, Tennessee, USA}
\address{$^2$Computer Science and Mathematics Division, Oak Ridge National Laboratory, Oak Ridge, Tennessee, USA}
\address{$^3$Bredesen Center for Interdisciplinary Research, University of Tennessee, Knoxville, Tennessee, USA}
\address{$^4$Electrical and Electronics Systems Research Division, Oak Ridge National Laboratory, Oak Ridge, Tennessee, USA}
\address{$^5$Center for Nanophase Materials Science, Oak Ridge National Laboratory, Oak Ridge, Tennessee, USA}
\address{$^6$Nuclear Security and Isotope Technology Division, Oak Ridge National Laboratory, Oak Ridge, Tennessee, USA}
\address{$^7$Computational Sciences and Engineering Division, Oak Ridge National Laboratory, Oak Ridge, Tennessee, USA}
\ead{humblets@ornl.gov}
\vspace{10pt}
\begin{indented}
\item[]Original 29 May 2016, Revised 3 August 2016
\end{indented}

\begin{abstract}
Developing devices that can reliably and accurately demonstrate the principles of superposition and entanglement is an on-going challenge for the quantum computing community. Modeling and simulation offer attractive means of testing early device designs and establishing expectations for operational performance. However, the complex integrated material systems required by quantum device designs are not captured by any single existing computational modeling method. We examine the development and analysis of a multi-staged computational workflow that can be used to design and characterize silicon donor qubit systems with modeling and simulation. Our approach integrates quantum computational chemistry calculations with electrostatic field solvers to perform detailed simulations of a phosphorus dopant in silicon. We show how atomistic details can be synthesized into an operational model for the logical gates that define quantum computation in this particular technology. The resulting computational workflow realizes a design tool for silicon donor qubits that can help verify and validate current and near-term experimental devices.
\end{abstract}

\vspace{2pc}


\section{Introduction}
Theory predicts quantum computing will speedup solutions to important problems in science, engineering, and security \cite{Nielsen2000}. Realizing these speedups will require a machine that is capable of executing computational gates on arrays of physical qubits over timescales that are much shorter than the influence of environmental noise and decoherence \cite{Metodi2006}. The fault-tolerant operation of quantum computers will require manipulating millions of qubits over billions of instruction cycles \cite{Ahsan2015}, while the architectures for these large-scale systems will require sophisticated execution and run-time control systems \cite{Britt2015}. However, the basic operating principles of quantum computation can be demonstrated using a more modest number of qubits. This includes demonstrating the non-local correlations inherent to entangled particles and showing the computational steps needed by few-qubit algorithms \cite{Monz2016,Debnath2016}. 
\par
Recent breakthroughs in the realization of donor qubits in silicon with high gate fidelities and long qubit coherence times have increased interest in these quantum computing architectures \cite{Pla2012, Pla2013, Muhonen2014}. However, the current proof-of-concept experimental demonstrations must be viewed as point solutions for the larger problem of designing robust quantum computing devices.  Key qualities of robust device operation include stability, reproducibility, and reliability, which are currently lacking in existing experimental systems. This is partly due to the uncertainty in the physical processes used for material fabrication, which hinders the experimental effort to realize specific physical designs. As insights into fabrication are gathered from early experiments, we may also expect modeling and simulation to provide guidance on how to design stable and reproducible qubit devices. Modeling and simulation have proven especially useful for the development of conventional CMOS (complementary metal-oxide semiconductor) processors, where TCAD (technology computer aided design) tools are standard for verifying designs generated from well characterized process models. Near-term progress in the development of robust quantum devices is poised to benefit from similar TCAD tools that capture the details of process variation, physical model uncertainty, and operational details in the computing context.
\par
\revised{
Currently, there is no single, standalone tool capable of capturing the quantum-mechanical, electrodynamical, and quantum computational features needed for the verification of silicon donor qubit device designs. This is because the multi-physics and multi-scale models required for describing such devices span across the simulation domains of existing state-of-the-art tool sets. Consequently, there is an outstanding need to integrate the physical and logical models for qubit devices into a single computational workflow that can provide assessments for expected performance against known limitations. In this contribution, we address the development of a computational workflow that can be used for modeling and simulation to verify silicon quantum computing devices.}  A qubit design tool is necessary to support efforts in fabricating quantum physical systems that express individual qubits, controlling their programmed interactions, and carrying out these interactions in a computationally meaningful way.  Our approach is based on a workflow framework that addresses the performance of small-scale doped silicon nanostructures, where models of physical layout and material design must be combined in order to evaluate the operating principles of an encoded qubit. This computational workflow exposes the ability to investigate key qualities of a device, e.g., gate fidelity, by tuning the design parameters. The computational models used here provide a natural bridge between efforts to fabricate and characterize silicon donor systems and efforts to program future multi-qubit systems \cite{Hill2015,OGorman2016}. 
\par
Our computational workflow complements state of the art experimental capabilities for investigating the development of silicon donor systems under different material and environmental conditions \cite{Jehl2016}. In addition, our approach to modeling and simulation mimics some of the long-standing methods used in CMOS device fabrication. This includes conventional TCAD tools, like Sentaurus Device from Synopsys, that provide robust methods for simulating the electrical, thermal, and optical properties of CMOS-based devices. These conventional TCAD tools are typically part of a larger simulation suite that intends to capture all aspects of conventional semiconductor chip development. However, the physical assumptions underlying these models are often limited to classical or semi-classical physics and, therefore, are not suitable for modeling strictly quantum effects. A recent example of a quantum TCAD tool was put forward by Gao et al., who developed QCAD as a TCAD device simulator that accounts for quantum effects in silicon quantum dot devices \cite{Gao2013, Gao2014}. The work of Gao et al.~not only demonstrates the need for specialized TCAD tools that address quantum computing devices, but also serves to clarify the diversity of quantum technologies available requiring such tools.
\par
The paper is organized as follows: We first present an overview of the silicon donor device model for quantum computing in Sec.~\ref{sec:si} followed by the development of a computational workflow to design silicon donor qubits in Sec.~\ref{sec:work}. We then describe a complete implementation of the design workflow in Sec.~\ref{sec:dev} with details about the electrostatic and quantum chemistry solvers in Sec.~\ref{sec:em} and Sec.~\ref{sec:qm}, respectively, and details about the gate operation model in Sec.~\ref{sec:gom}. Our final remarks are presented in Sec.~\ref{sec:con}.
%%%%%%%%%%%%%%%%%%%%%%%%%%%%%%%%%%%%%%%%%%%%%%%%%%%%%
\section{Silicon Donor Qubit Models}
\label{sec:si}
Among the many different approaches to qubit fabrication, donors in silicon represent a promising path for encoding and manipulating qubits of information \cite{Zwanenburg2013}. As originally proposed by Kane, the idea is to encode quantum information into the nuclear spin state of a dopant atom like phosphorous ($^{31}$P) that is shallowly embedded in a silicon lattice composed of $^{28}$Si \cite{Kane1998}. Because this silicon isotope has zero nuclear spin, the spin-1/2 phosphorous atom is ideally isolated from other spin contaminants \cite{itoh2014mrs}. Several works that followed the Kane proposal have also outlined how to realize qubits using the unpaired donor-bound electron, opening up the potential for easier addressability and faster gate operation \cite{Hill2015,vrijen00pra, Hillprb2005,  Hollenberg2006prb}.  Recent experiments using isotopically purified $^{28}$Si have demonstrated single-qubit coherence times ($T_2$) over 30 seconds (nuclear spin) and 0.5 seconds (electron spin) at cryogenic temperatures \cite{Muhonen2014}. 
\par
A silicon donor qubit can be electrically manipulated by addressing the donor electron wave function with external gate electrodes. The donor electron wave function serves as a `handle' for addressing the spin states through the hyperfine interaction, which is directly proportional to the spin density at the donor site \cite{Feher1959pr}. The spin states of neighboring donors are coupled via the electron exchange interaction, which can be controlled by an external potential. Recent experiments have realized basic controls for phosphorus qubits in silicon (Si:P), however, a multi-qubit testbed is yet to be realized \cite{Pla2012,Pla2013,Muhonen2014,Pica2014,Morello2010,Kalra2014,Laucht2015,Dehollain2016}.
\par
A key requirement for donor-based quantum computing architectures is the manipulation of individual donor electron wave functions over well-defined regions of space and time in the presence of material variations, finite temperatures, environmental noise, and electronic fluctuations \cite{Zwanenburg2013}. The donor electron wave function oscillates several times across the interaction region between qubits, due to interference effects arising from the degeneracy of the 6-valleys in the conduction band of silicon. The exchange coupling ($J$) between donors is sensitive to these oscillations, which occur with a periodicity on the order of the silicon lattice constant \cite{Koiller2002prl}. Uncertainty in the position of the donor atoms therefore leads to unpredictable exchange couplings. Given the variability in the placement of the donors \cite{van2015jpcm}, the external potential must be tuned sufficiently to obtain the desired value of $J$. In addition, the donor position needs to be well-defined relative to the surface electrodes that lie along an atomically rough boundary. These process errors complicate control of the resulting qubit. Moreover, they are in addition to the decoherence caused by external noise, defects, spectral diffusion, and charge buildup at the electrode interfaces, which undermines the desired dynamics \cite{Calderon2007}.
\par
Device sensitivity to the electronic structure details imposes a formidable challenge to fabricating silicon donor qubits. But this challenge is not insurmountable as evidenced by increasing precision in sample fabrication \cite{Fuechsle2012nn}, reduced decoherence from isotopic purification of the silicon substrate, and recent efforts to directly visualize the donor electronic state \cite{salfi2014nm}. Very recently, Laucht et al.~realized a single-qubit device from a lone $^{31}$P atom embedded in isotopically pure $^{28}$Si \cite{Laucht2015}. In particular, the electron and nuclear spin states of the P donor were manipulated using nanoscale gates to perform single-qubit logic gates and measure experimental signatures of coherence. Accurate characterization of the device in Ref.~\cite{Laucht2015} required precise knowledge about the donor, including parameters such as the exact donor position, material strain and electrostatics in the vicinity of the donor. This demonstration underscores the need to better understand silicon donor physics at both a material and design level in order to improve device behavior.
%%%%%%%%%%%%%%%%%%%%%%%%%%%%%%%%%%%%%%%%%%%%%%%%%%%%%
\section{Computational Workflow}
\label{sec:work}
The sensitivity of silicon donor physics to atomistic details, including single-atom defects, charge noise, and inter-donor spacing, emphasizes the significance of using microscopic models to guide device development. However, these models challenge existing TCAD tools because they require material models and simulation techniques that incorporate quantum physics. In addition, device models for silicon donor qubits are sufficiently distinct from existing CMOS-based transistors and integrated circuits as to warrant a dedicated design framework. In this section, we develop a computational work flow sufficient to express the behavior and function of state of the art silicon donor devices.
\par
\begin{figure}[t]
\centering
\includegraphics[width=1.0\columnwidth]{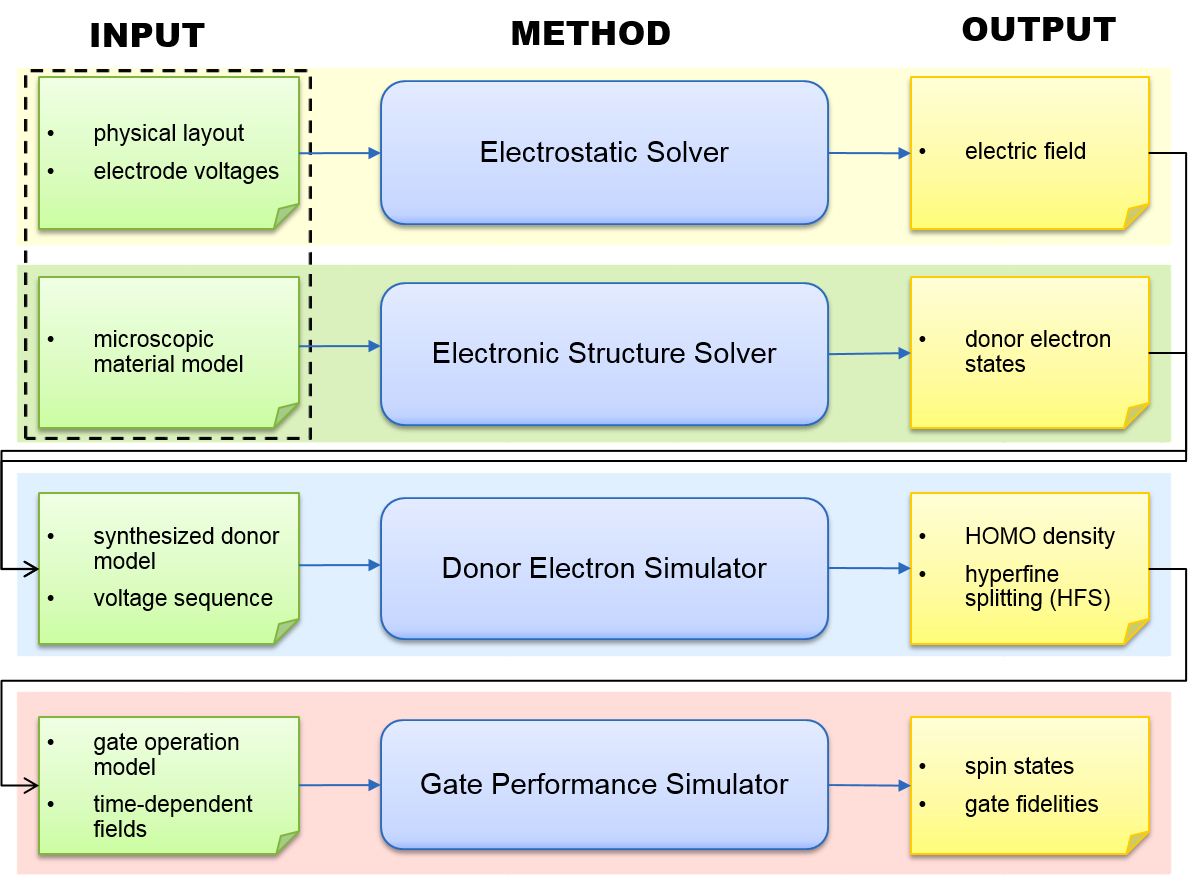}
\caption{The computational workflow to simulate gate performance for a silicon donor qubit. Inputs models defining the physical layout and material composition are simulated to generate intermediate representations of the donor electron wave function. These intermediate models are synthesized to define a gate operation model that is simulated to calculate the nuclear and electronic spin states and resulting gate fidelity.}
\label{fig:workflow}
\end{figure}

Our computational workflow for designing silicon donor qubits makes use of a multi-stage model that synthesizes intermediate solutions into a complete device representation. An overview of this workflow is shown in Fig.~\ref{fig:workflow}, in which electromagnetic and material input models are first simulated to recover the electrostatic field and electronic structure, respectively. These results are synthesized into a field-dependent model of the donor electron wave function. The synthesized model represents a silicon donor qubit design parameterized by the value of the electrostatic field. In the next stage, we use this model to simulate changes in the donor electron wave function induced by varying the applied electric field. The results of these simulations provide the necessary atomic details to describe the donor electron wave function, the hyperfine splitting, and electron exchange interaction. The final state of the workflow uses these atomic details to construct a gate operation model that describes the behavior of the silicon qubits driven by specific voltage (field) sequences. The output of the gate operation simulation provides the time-dependent electronic and nuclear spin states induced by the applied electric field. In particular, these spin states describe the expected gate fidelities and estimate the system decoherence time. 
\par
The required input to the workflow is a physical device layout defining the electronic circuit geometry, voltages, sources and drains as well as a partitioning of different material regions in the device. These material regions define the physical properties needed to calculate the electric field that exists at the donor location. An example physical layout is shown in Fig.~\ref{fig:comsol_device}, which is a model of the device used in recent experimental demonstrations by Laucht et al.~\cite{Laucht2015}. Material properties for the surfaces and layers of this device must be specified to facilitate subsequent simulation of the electric field near the donor system. The specification also includes the operating temperature for the device, which may be on the order of sub-Kelvin temperatures for silicon donor qubits.
%%%%%%%%%%%%%%%%%%%%%%%%%%%%%%%%%%%%%%%%%
%%%%%%%%%%%%%%%%%%%%%%%%%%%%%%%%%%%%%%%%%
\par
In addition to the electromagnetic properties of the layout, the workflow also requires a microscopic description of the silicon lattice in the region of the donor atom. This is necessary to simulate the electronic state of the donor atom. Simulations of the electron wave function may make use of different levels of theory, and these choices will impact the format by which the material model should be presented. A common format is the crystal geometry of the sample materials including the location of the donor atom. Any input must also account for the relative position of the donor with respect to the electrode layout including its material layers. For example, the position of the donor atom relative to the surface of the device determines the amount of strain present in the silicon due to the lattice mismatch of the substrate with the silicon dioxide and metal electrode layers.
\par
An important consideration for current qubit development efforts is understanding how variations in material properties, environmental conditions, and electrode design impact the utility of silicon donor qubits for quantum logic operations. In particular, donor species and placement as well as applied fields influence the behavior of the hyperfine coupling, the spin states and the resulting gate fidelities. Consequently, the input models to our computational workflow may have a probabilisitic component that defines a priori distributions for the fluctuating physical parameters. For example, the relative position of the donor atom may be known with limited accuracy, as found in recent metrological studies \cite{Mohiyaddin2013nl}. The variability of positions can be accounted for by considering a statistical distribution, and the results may then be statistically combined to obtain an average fidelity.
%%%%%%%%%%%%%%%%%%%%%%%%%%%%%%%%%%%%%%%%%%
\section{Qubit Design}
\label{sec:dev}
As shown in Fig.~\ref{fig:workflow}, the electric and material input models pass through initial simulation stages that generate, respectively, the electric field defined across the entire device and the electron orbitals for the unperturbed donor atom in the presence of material defects. The methods that are available for these simulations steps are varied, and we describe the expectations for the electromagnetic and material simulations in Sec.~\ref{sec:em} and Sec.~\ref{sec:qm}, respectively. In our workflow, the output from these models are synthesized together by first extracting the electric potential near the donor atom in the material sample. In Sec. ~\ref{sec:gom}, we describe how this model can be used to simulate gate operation with atomistic detail. 
%%%%%%%%%%%%%%%%%%%%%%%%%%%%%%%%
\subsection{Electrostatic Field Simulations}
\label{sec:em}
\begin{figure}[ht]
\centering
\includegraphics[width=1\columnwidth]{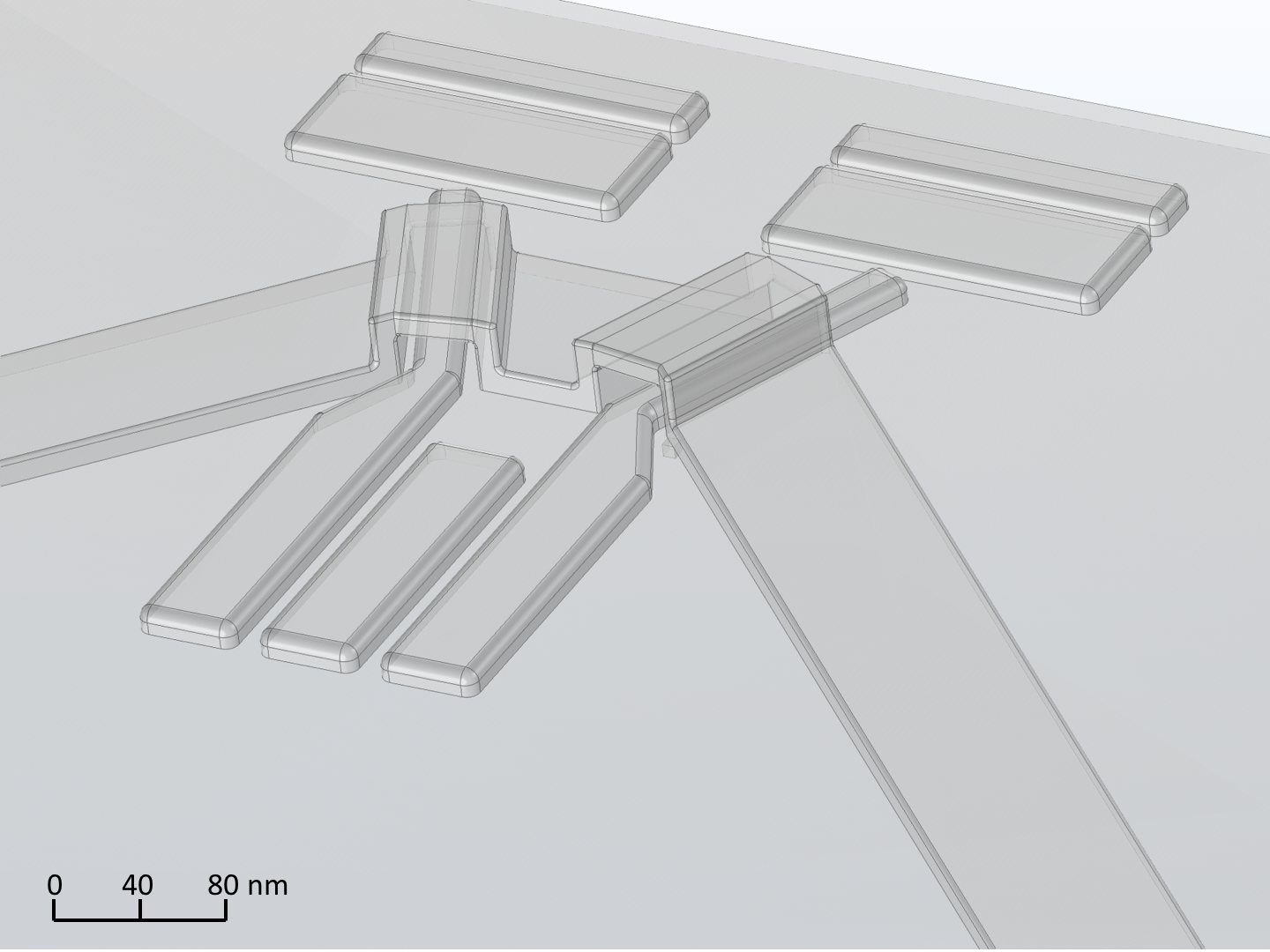}
\caption{A physical layout representing the model electrodes and underlying substrate. The model presented here is based on the recent device used in Ref.~\cite{Laucht2015}}
\label{fig:comsol_device}
\end{figure}

The first stage of the workflow requires a specification of the physical layout of the device. An example of this specification is shown in Fig.~\ref{fig:comsol_device}, where a physical layout based on the experimental device tested by Laucht et al.~is presented ~\cite{Laucht2015}. This particular device consists of multiple DC and AC electrodes used to tailor the electromagnetic environment of the donor atom embedded in the underlying silicon lattice. In general, the layout must account for the bulk material properties of the aluminum electrodes, the $\textrm{SiO}_2$ interface layer and any intrinsic charge regions. This requires assigning material properties to each region of the layout, including the silicon, silicon oxide, aluminium, and vacuum regions. In addition, the input specification must also assign voltages to each (DC) electrode. Although the source and drain regions are not visible in Fig.~\ref{fig:comsol_device}, they are included in the model as well as additional electrodes are connected to ground the substrate. 
%%%%%%%%%%%%%%%%%%%%%%%%%%%%%%%%%%%%%%%%%
%%%%%%%%%%%%%%%%%%%%%%%%%%%%%%%%%%%%%%%%%
\par
\begin{figure}[ht]
\centering
\includegraphics[width=\columnwidth]{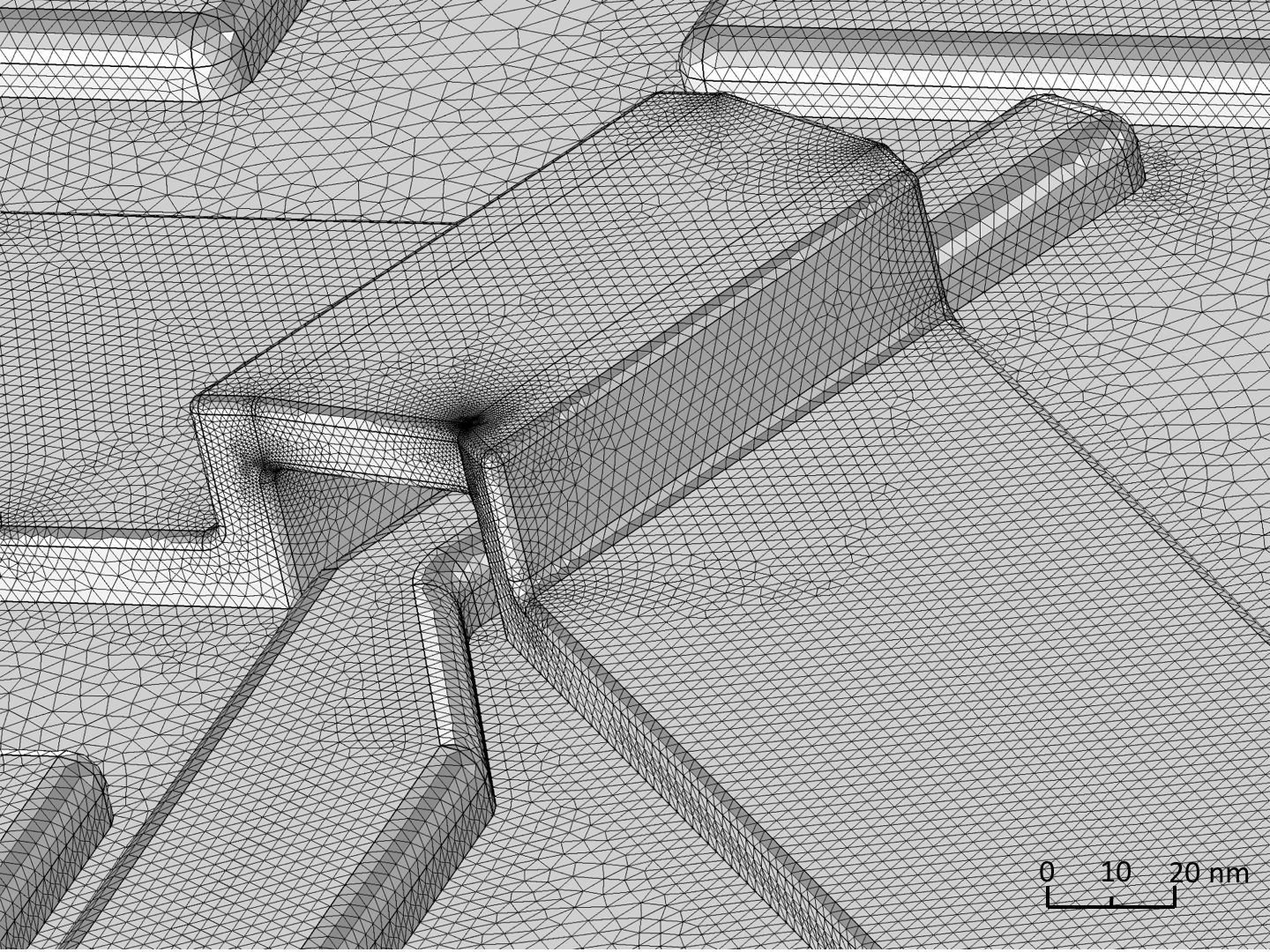}
\caption{A view of the mesh used for simulating the electrostatic field of the device in Fig.~\ref{fig:comsol_device}. The mesh shown resulted in ~22 million elements.}
\label{fig:comsol_mesh}
\end{figure}

Simulating the electromagnetic field corresponding to the physical layout requires solving Poisson's equation over the entire device model. This is typically accomplished by first meshing the entire device with a resolution sufficient to accurately represent changes in the underlying potential. As shown in Fig.~\ref{fig:comsol_mesh}, meshing defines the points at which the potential is calculated and these points must be of sufficient density such that the boundary conditions between different materials satisfy the electromagnetic field equations. The process of finding the optimal mesh depends on the geometry of the device, although many implementations of electrostatic solvers provide options for automated mesh generation. For the physical layout shown in Fig.~\ref{fig:comsol_device}, we obtained converged solutions for the field near the location of the donor using an extremely fine mesh created with mesh spacing restricted to 0.1~nm for a 3x3x3~nm${}^3$ block.
%%%%%%%%%%%%%%%%%%%%%%%%%%%%%%%%%%%%%%%%%
%%%%%%%%%%%%%%%%%%%%%%%%%%%%%%%%%%%%%%%%%
\par
\begin{figure}[ht]
\centering
\includegraphics[width=1.05\columnwidth]{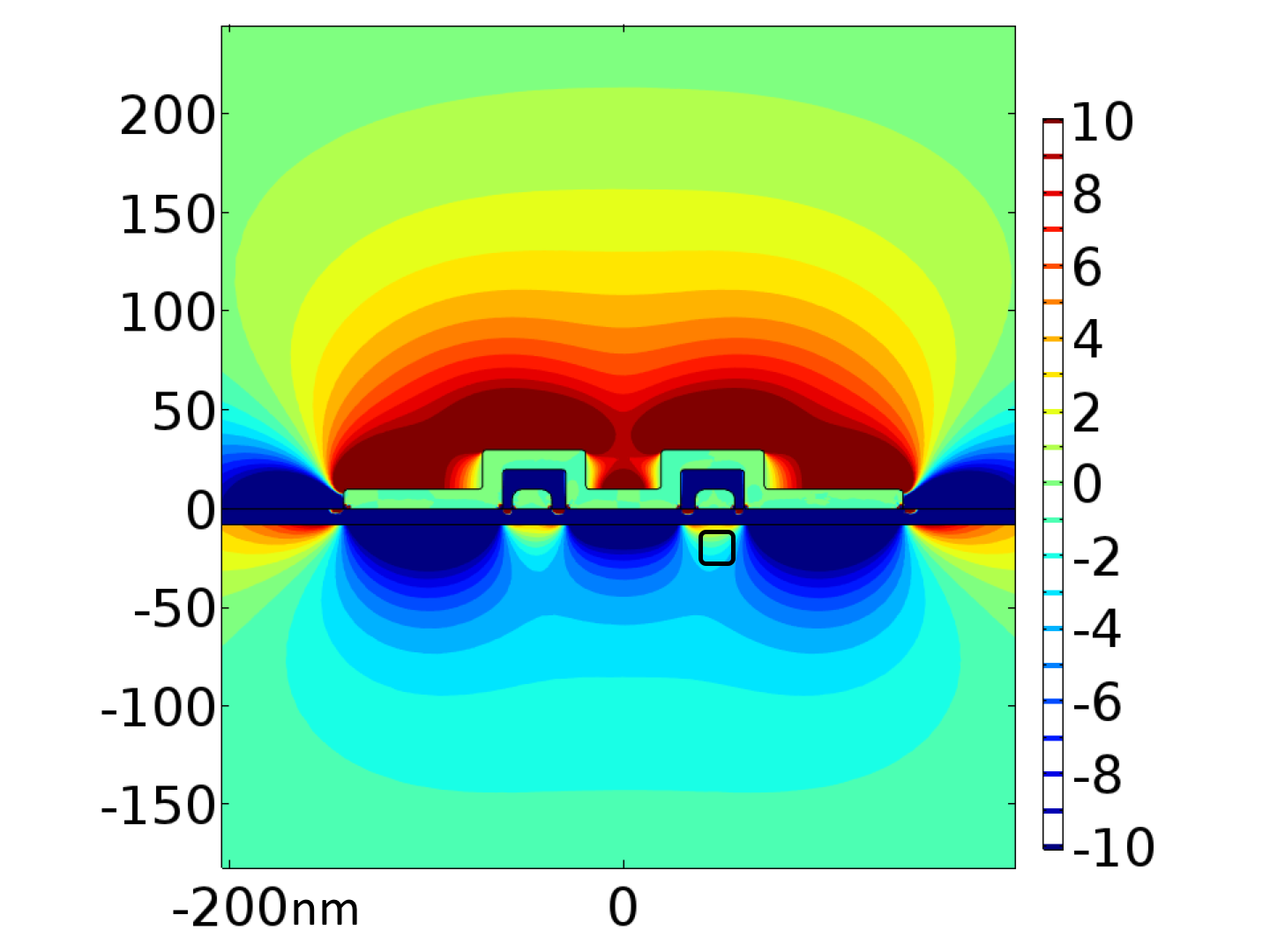}
\caption{A simulation of the electric field for the device shown in Fig.~\ref{fig:comsol_device}. This plot shows the value of the field in the direction vertical to the plane of the device measured in V/$\mu$m. We use values of the simulated field taken from the highlighted square located under the right electrode. This corresponds with is the estimated location of the phosphorus atom and the area where increased resolution meshing is desired.}
\label{fig:comsol_electric_field}
\end{figure}

We have tested several commercial solvers for solving the electrostatic field of the prototype device given in Fig.~\ref{fig:comsol_mesh}. We have found the commercial electrostatic solver packages from COMSOL to be capable of modeling and simulating the electrostatics of these complex nanostructure geometries. An example of the resulting electrostatic field produced is presented in Fig.~\ref{fig:comsol_electric_field}, which demonstrates the degree of expected field variability across a mesoscopic slice of the device. Regions in close proximity to the addressing electrodes can have relatively large electric field derivatives. These regions are of specific interest for controlling the electron wave function of the donor atom.
\par
However, we anticipate that a challenge for these types of electrostatic field simulations is the degree of meshing needed. Whereas the solution to the electrostatic fields depends on satisfying boundary conditions over the complete nanoscale structure, there is also a requirement to provide sub-nanometer spatial resolution in the region of the donor. This is because the best basis functions for calculating the donor electron wave function will typically represent well-localized atomic orbitals. Because these orbitals are defined on much smaller length scales than the rest of the device, high density meshes will be needed in regions close to the donor location. Adaptive meshing alleviate some of the concern from increasing size, but when the location of the donor is variable, repeated simulations with different meshing will become a significant computation. However, sampling over the distribution of donor location will be an important measure of donor sensitivity to electrode design.
%%%%%%%%%%%%%%%%%%%%%%%%%%%%%%%%%%%%%%%%%
%%%%%%%%%%%%%%%%%%%%%%%%%%%%%%%%%%%%%%%%%
%%%%%%%%%%%%%%%%%%%%%%%%%%%%%%%%%%%%%%%%%
\subsection{Donor Electron Wave Function Simulations}
\label{sec:qm}
Quantum mechanical modeling of the silicon lattice and the embedded donor require solving for the electronic structure of the combined system. In particular, the electronic structure of the donor electron plays a bridging role that defines the effect of the applied electromagnetic fields on the time-dependent dynamics of the donor spin states. We now discuss the challenges related to electronic structure simulation for the donor electron wave function, especially with respect to the computational complexity and desired accuracy needed for subsequently describing gate operation.
\par
A main challenge in the electronic structure simulation of the electron donor state is the computational complexity of the underlying methods. For example, density functional theory (DFT) is a workhorse of material science modeling that typically operates within the 100 to 1,000 atom regime on standalone workstations while implementations leveraging massively parallel computing resources can reach up to 10,000 atoms. A 5~nm wide silicon crystal cube consists of 10,000 atoms and 140,000 electrons. This volume is relatively small compared to the physical layout in Fig.~\ref{fig:comsol_mesh} but represents the high end of what is possible for DFT calculations. The root cause behind this practical limitation is that the computational cost for DFT typically scales cubically with system size due to the required linear algebra matrix-matrix operations. 
\par
An example of the scaling expected for these DFT calculations is shown in Table \ref{tab:dft-cost}. Our example uses the NWChem suite \cite{nwchem} to calculate the electronic structure for Si:P nanocrystals up to 3~nm in size. Using Becke's three-parameter hybrid functional in combination with Lee-Yang-Parr correlation functional (B3LYP) \cite{becke,LYP}, we performed tests of DFT scaling with various basis sets and degrees of computational parallelism. We typically used between 10 and 100 basis functions per atom, where the total number of basis functions determines the size of the Hamiltonian matrix. The electronic structure of the donor is then solved by iteration in a self-consistent field (SCF) process through diagonalization of the Hamiltonian. Typically up to 50 iterations (diagonalizations) were required. Our test used parallelized computations with up to 72 CPU nodes, where each CPU node has 2 Haswell processors (thus 24 cores per node).  As shown in Table \ref{tab:dft-cost}, over 1,728 CPU cores were used for several hours to calculate the electronic structure energy for the largest Si:) nanocrystal considered here (diameter 3~nm).  
%%%%%%%%%%%%%%%%%%%%%%%%%%%%%%%%%%%%%%%%%%%%%%%%%%
\begin{table*}
{\footnotesize
\caption{Computational time required for DFT calculations of Si:P nanocrystals.  A service unit (SU) equals the number of cores multiplied by the number of hours.}
\label{tab:dft-cost}
%\begin{tabular}{rrcrrrrr}
\begin{tabular}{rrcrrrrrr}
\hline\hline
Sample   & Number   & Basis set   &  Number   &  Number   & Number   &  Seconds per    & CPU hours \\  
diameter & of atoms &   (P/Si/H)  & of basis  &  of nodes & of cores &  iteration      & per iteration  \\
         &          &             & functions &           &          & (Walltime)      & (1SU=1hour) \\
\hline\hline
1 nm  &   65  &   3-21++g*/3-21++g*/3-21G       &    739 &    4 &   96 &    22 sec  &   0.59 SU \\
1 nm  &  65   & aug-cc-pvtz/aug-cc-pvdz/cc-pvdz &  1051  &    4 &   96 &   400 sec  &  10.7 SU \\
1 nm  &  65   & aug-cc-pvtz/aug-cc-pvdz/cc-pvdz &  1051  &    6 & 144  &   291 sec  &  11.6 SU \\
\hline
2 nm & 295  &   3-21++g*/3-21++g*/3-21G           &   4265 &    1 &    24 &  33,000 sec  & 220 SU \\
2 nm &  295 &  3-21++g*/3-21++g*/3-21G            &   4265 &    4 &    96 &   7809  sec  & 208 SU \\
2 nm &  295 &  3-21++g*/3-21++g*/3-21G            &   4265 &   16 &   384 &   1004  sec  &  107 SU \\
\hline
3 nm & 1005 &  3-21++g*/3-21g/sto-3g              &   9475 &   72 &  1728 &  1427   sec  & 685 SU \\
\hline\hline
\end{tabular}
}
\end{table*}
%%%%%%%%%%%%%%%%%%%%%%%%%%%%%%%%%%%%%%%%%%%%%%%%%%
\par
As expected, DFT calculations are not tractable with increasing system size. The development of models with reduced complexity for electronic structure calculations is therefore necessary for applying computational chemistry to models of silicon donor devices. One approach for reducing the computational complexity is to simplify the theory by neglecting core electrons that are typically not involved in chemical bonding. Another simplification is to leverage the locality of interactions by ignoring integrals between well-separated orbitals. Tight-binding (TB) \cite{nemo-I,nemo-II} and density-functional tight binding (DFTB) \cite{dftb} methods that make use of these approximations are computationally attractive alternatives for electronic structure modeling of silicon donor qubit devices. Whereas TB methods have been used extensively for modeling silicon donor systems \cite{salfi2014nm,klimeck,Rahman2009,Ahmed2009,Mohiyaddin2014,Wang2016}, DFTB methods have not yet been extensively applied.
\par
\revised{
The relative error acquired by adopting different theoretical models for solving the electronic structure of the donor atom play an important role in simulating silicon qubit device physics. For example, a central physical quantity in the model of the donor atom as a qubit is the value of the hyperfine splitting (HFS). In principle, the HFS is the leading term in the nuclear spin-electronic interaction for the Si:P qubit. The isotropic HFS, also known as the Fermi contact interaction, is proportional to the spin density on the donor atom, where the spin density is defined as the difference between the densities for the spin up ($\rho_{\alpha}(r)$) and spin down ($\rho_{\beta}(r)$) electronic states. However, this difference is often approximated by the density of the unpaired electron $|\phi(r)|^2$ that is bound to the donor atom. Therefore, the isotropic HFS is given as
\begin{equation}
A_{\mathrm{iso}} = -\frac{8}{3} \pi \langle \mu_n \cdot \mu_e \rangle  \left[\rho_\alpha(r_0) -\rho_\beta(r_0)  \right]
            \approx -\frac{8}{3} \pi \langle \mu_n \cdot \mu_e \rangle  | \phi (r_0) |^2  
\label{eq:Aiso}
\end{equation}
where $\mu_n$ is the nuclear magnetic moment, $\mu_e$ is the electron magnetic moment, and $r_0$ is the donor position. 
}
\par
\begin{figure}[t]
\begin{tabular}{ccc}
\includegraphics[width=0.3\columnwidth]{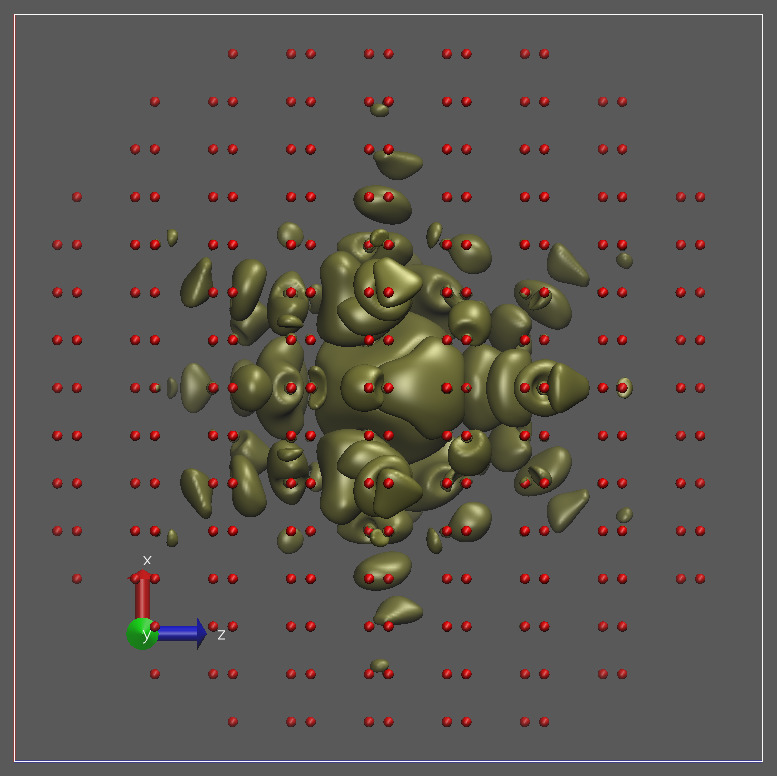} &
\includegraphics[width=0.3\columnwidth]{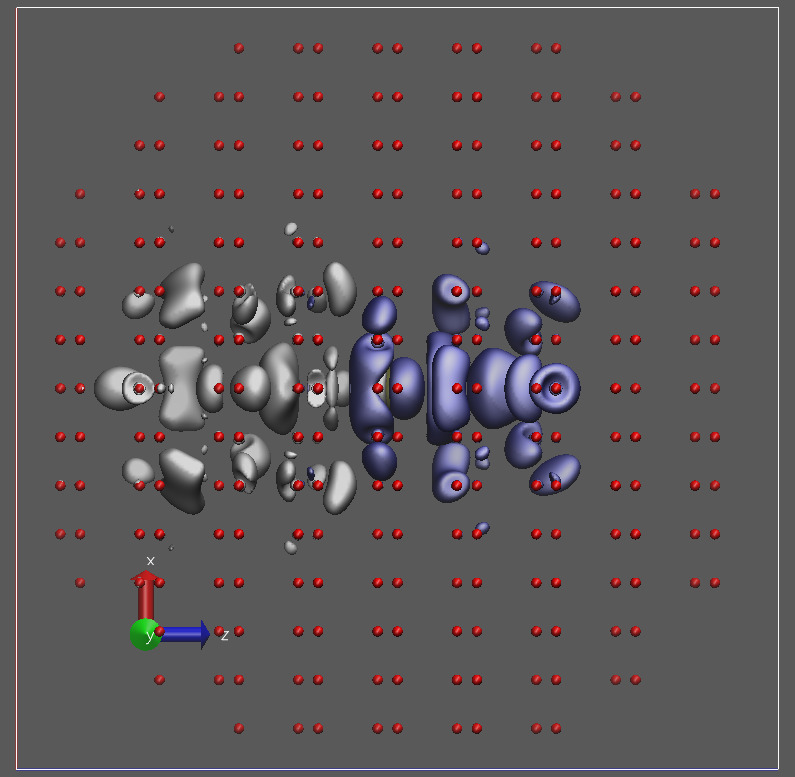}  &   
\includegraphics[width=0.4\columnwidth]{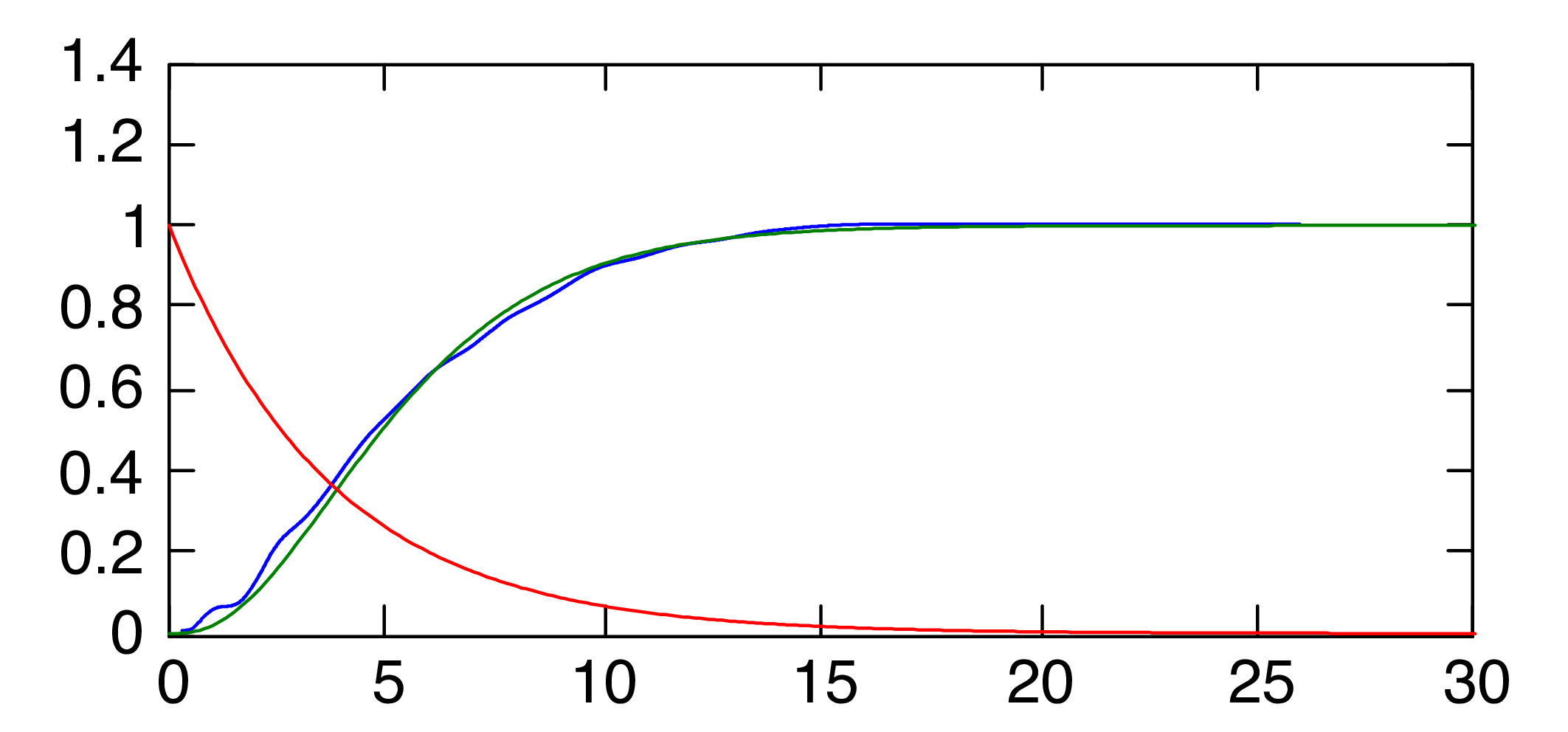} \\
(a) & (b)  & (c) \\
\end{tabular}
\caption{Electronic structures of phosphorous doped silicon nanocrystal Si:P with a size of 3~nm. Silicon centers are shown as red dots and the phosphorus atom is at the center of the nanocrystal. (a) Isosurface for the HOMO in the absence of applying an electric field. (b) Differential distribution of the HOMO under an electric field along the horizontal  axis. (c)  Cumulative radial distribution from  DFT calculations (blue)  and its fit  to Slater type orbital  (green). Corresponding radial distribution is shown in red  as a function  of distance (in Angstrom).}
\label{fig:HOMO}
\end{figure}

\revised{
As an example, Fig.~\ref{fig:HOMO}(a) shows the highest occupied molecular orbital $|\phi(r)|^2$ for a Si:P nanocrystal  obtained using DFT simulations and Fig.~\ref{fig:HOMO}(b) shows the influence of an applied electric field on the differential distribution. An analysis of the HOMO density indicates that the radial distribution of the electron orbital closely matches that of a hydrogenic S type function that can be represented by a Slater-type orbital. The observed agreement is in contrast to the Gaussian type orbital expected for a quantum dot bound by a harmonic potential. In particular, the HOMO density decays very rapidly with respect to the distance from the donor, as shown in Fig.~\ref{fig:HOMO}(c). These results suggest that  although the shape of the orbital for the unpaired electron is complicated, it may be acceptable to use approximate the spin density by a much simpler, analytic form. This approximation can greatly simplify the electronic structure calculations for the donor atom in the presence of variable applied fields. For example, using a Slater-type orbital as an approximation to the actual donor electron orbital provides a convenient method for calculating how the HFS in Eq.~(\ref{eq:Aiso}) scales with applied electric field. In lieu of recomputing the exact electron orbital, which may be time consuming, a perturbation to the approximate analytical orbital can be used to estimate the corresponding HFS. The validity of this perturbative approach depends on the material environment and symmetry of the applied field. This approximation is most likely to be valid when the environment is highly symmetric.
}
\par
\begin{figure}[t!]
\centering
\includegraphics[width=0.5\columnwidth]{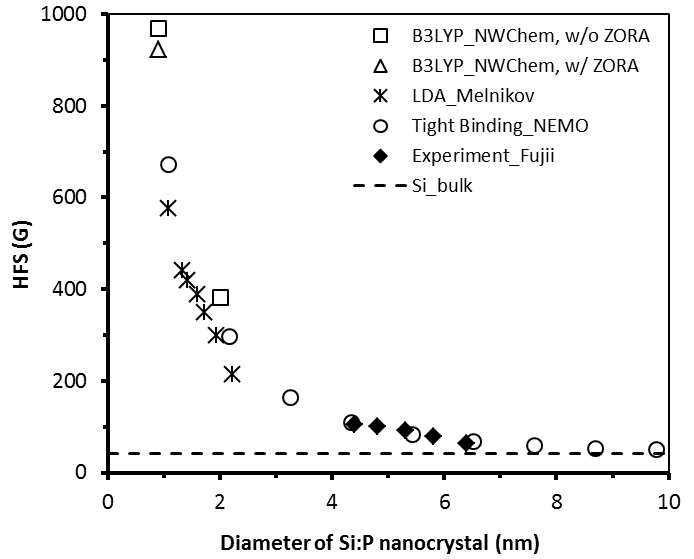}
\caption{Hyperfine splittings (HFS) in Gauss as a function of the diameter of Si:P nanocrystals calculated at different levels of theory compared to experimental data. The tight binding calculations using NEMO 3-D were performed for nanocrystals up to 30 nm but only the results from 1 to 10 nm are shown.}
\label{fig:hfc}
\end{figure}

\revised{
We now address how the theoretical difference between the spin density and the HOMO density may be handled and what trade-off may be gained between the accuracy of adopting an approximate orbital. The answers to these questions are obtained through the comparison of theoretically calculations and experimental results. As a specific example, we compare calculations of the HFS for Si:P nanocrystals using various theoretical methods against experimental measurements of the same quantity, as shown in Fig.~\ref{fig:hfc}. The experimental HFS value for P in bulk Si is known to be 42 G, while experimental values for Si:P nanocrystals embedded in insulating phosphosilicate glass matrices were previoulsy measured by Fujii et al. using electronic spin resonance (ESR) \cite{Fujii2002}. In those experiments, the Si:P nanocrystals have a diameter $d$ in the range of 4.4-6.4~nm with a standard deviation of $\sim$1~nm. They were found to have a small number of deep dangling bond defects at the Si-SiO$_2$ interfaces which are filled by P doping as indicated by the increasing photoluminescence intensity at a low P concentration range and by the infrared optical absorption beyond a low P concentration threshold. 
}
%%%%%%%%%%%%%%%%%%%%%%%%%%%%%%%%%%%%%%%%%%%%%%%%%%
%%%%%%%%%%%%%%%%%%%%%%%%%%%%%%%%%%%%%%%%%%%%%%%%%%
\par 
\revised{
We make our comparison using several theoretical models that differ in accuracy and computational cost, as shown in Fig.~\ref{fig:hfc}. Hhybrid DFT calculations with B3LYP are very expensive, cf. Table \ref{tab:dft-cost}, especially when the relativistic zeroth-order regular approximation (ZORA) is included \cite{zora1,zora2}. DFT calculations within the local density approximation (LDA) performed by Melnikov and Chelikowsky \cite{chelikowsky} are expected to be computational more efficient due to the lack of exact Hartree-Fock exchange and the employment of pseudopotentials. These \textit{ab initio} DFT methods can easily be afforded for the smallest nanocrystals with $d$ in the range of 1-2~nm, with the nanocrystals modeled by spherical Si nanoparticles with H terminations on the surface to passivate the dangling bonds. However, for even larger diameters, the computational costs of DFT were too demanding. However, we found that the semi-empirical TB models implemented within the NanoElectronic Modeling in three dimensions (NEMO 3-D) code were capable of addressing these larger sized particles \cite{nemo-I,nemo-II}. Using NEMO-3D, the nanocrystals were modeled by a cubic box with a zinc blende lattice of Si and a cube length of $a$ in the range of 1-30~nm. The surface was not H-terminated but rather passivated by shifting the energy of the dangling bonds by 30 eV \cite{nemo-ug}. The Si atom in the center of the particles was replaced with a dopant P atom giving Si:P nanocrystals of different sizes. The geometries were not relaxed because it has been shown in the literature that there is no significant relaxation of the surrounding Si atoms when P serves as the dopant in a Si lattice \cite{chelikowsky}. However, doping with other group V elements such as As, Sb, or Bi may render a large geometrical distortion \cite{Usman2015}. }
%%%%%%%%%%%%%%%%%%%%%%%%%%%%%%%%%%%%%%%%%
%%%%%%%%%%%%%%%%%%%%%%%%%%%%%%%%%%%%%%%%%
%%%%%%%%%%%%%%%%%%%%%%%%%%%%%%%%%%%%%%%%%%
\par
\revised{
For the \textit{ab initio} DFT calculations, the isotropic HFS values were calculated by the Fermi contact interaction between the electron spin and the nuclear spin at the dopant P using Eq.~(\ref{eq:Aiso}). For the semi-empirical TB method, the HFS was not explicitly calculated but instead found using a scaling method given a known HFS value for bulk Si as
\begin{equation}
\frac{A_{\mathrm{iso}}(d)}{A_{\mathrm{iso}}(\mathrm{bulk})}  = \frac{|\phi(d,r_0)|^2}{ |\phi(\mathrm{bulk},r_0)|^2}  
\label{eq:Aiso_scale}
\end{equation}
where $A_{\mathrm{iso}}(\mathrm{bulk}) = 42$ G and we assume the bulk configuration corresponds with $a = 30$~nm. This scaling method is also used in the literature to study the Stark shift of the HFS in Si:P devices \cite{klimeck,martins}
\begin{equation}
\frac{A_{\mathrm{iso}}(E)}{A_{\mathrm{iso}}(0)}  = \frac{|\phi(E,r_0)|^2}{ |\phi(0,r_0)|^2}  
\end{equation}
where $E$ is the electric field. As can be seen from Fig.~\ref{fig:hfc}, the experimental HFS values for Si:P nanocrystals with $d$ in the range of 4.4-6.4~nm are 1.5-2.5 times larger than the bulk value and display a trend of increasing HFS with decreasing $d$. This strong particle size dependence was attributed to the quantum confinement of the unpaired electron spin in the nanocrystal. This trend is further verified by the \textit{ab initio} DFT calculations of Melnikov and Chelikowsky using the LDA functional \cite{chelikowsky}. Our own B3LYP calculations for $d$=1-2~nm nanocrystals using Eq.~(\ref{eq:Aiso}) gave the same trend, albeit with larger HFS values than the LDA results. For the smallest 0.9~nm nanocrystal, the relativistic ZORA calculation gave a smaller HFS than that without inclusion of relativistic effects. It is thus expected that with ZORA, the HFS for the 2~nm nanocrystal calculated by B3LYP could be closer to the LDA results. Note that Dunning's correlation consistent Gaussian basis set of cc-pVDZ was adopted for Si, P, and H for the smallest 0.9~nm nanocrystal but it was reduced to 3-21+G* for Si and P and 3-21G for H for the larger 2~nm nanocrystal. Studies for the smallest nanocrystal using different Gaussian basis sets indicate 20\% of changes for HFS values. Nevertheless, the results still show a trend of increasing HFS with decreasing nanocrystal size $d$. In comparison, the semi-empirical TB calculations by the scaling method shown in Eq.~(\ref{eq:Aiso_scale}) reproduced the HFS values and the trend quite well. However, if Eq.~(\ref{eq:Aiso}) is used to replace the scaling method, the HFS values would be lower than the bulk HFS value of 42 G beyond a particle size of 3~nm, due to the underestimated spin densities at the P nucleus.
}
\par
\revised{
In addition, in the \textit{ab initio} DFT calculations, spin-unrestricted open-shell wave functions were employed by giving spin-up and spin-down electrons different spatial orbitals. Consequently, the electron density $|\phi(r_0)|^2$ for the HOMO at the P nucleus are smaller than the spin density $(\rho_{\alpha}(r_0) - \rho_{\beta}(r_0))$ at the P nucleus by 10\%. By comparison, the TB calculations were performed using spin-restricted open-shell wave functions and therefore the HOMO densities are exactly the same as the spin densities at the P nucleus. We expect that the HOMO densities in the tight binding calculations are good approximations of the real spin densities due to the small changes from the spin densities as can be seen from the \textit{ab initio} DFT calculations. In short, the HFS values were found to depend on the level of theory, the size of the basis sets, and the spin restriction. Quantitative theoretical predictions require rigorous validations of theory which will be the focus of our future work.
}
\par
\revised{
A key challenge for fabricating these devices is controlling the location of the donor atom. In addition, there are no characterization methods that can pinpoint the location of the donor after device fabrication. State-of-the-art efforts reduce the uncertainty in the donor location using indirect position measurements, which effectively perform triangulation to narrow down the possible locations of the donor \cite{Mohiyaddin2013nl}. However, there are recent results that suggest the HFS at these locations may not be necessarily uncorrelated. In particular, the electron orbital wave function and HFS may have a weak dependence on donor translations along a plane parallel to the $\mathrm{Si/SiO_2}$ interface \cite{Mohiyaddin2014}. By contrast, translations in the depth of the donor show an approximately exponential dependence for the HFS \cite{Mohiyaddin2014}. These relatively simple models for the dependence of the donor wave function and the associated hyperfine splitting on donor position suggests that sampling over the distribution of positions can be made efficient, thereby reducing the computational cost considerably.
}
%%%%%%%%%%%%%%%%%%%%%%%%%%%%%%%%%%%%%%%%%
%%%%%%%%%%%%%%%%%%%%%%%%%%%%%%%%%%%%%%%%%%
\subsection{Gate Operation Model}
\label{sec:gom}
We now describe how the electrostatic and material models for silicon donor devices can be synthesized to generate a gate model describing how the nuclear and electron spin states of the donor system evolve in the presence of time-dependent changes to applied bias. In particular, we intend to define how this new model accounts for the instantaneous state of the donor and its changes in time. The instantaneous state is a fundamental description of the qubits represented by the electron and nuclear spin states that, within the context of quantum physics, provides all knowable information about the system. Therefore, we present a framework by which the complete information of the donor can be derived from the earlier workflow stages.
\par
Our use case for this stage of the workflow is to simulate the behavior of a device whose physical layout and composition served as input to the previous stages. Gates within this single-donor model correspond to well-characterized (continuous) sequences of the applied bias that tune the hyperfine and Zeeman level spacings. An externally applied AC magnetic field then induces transitions between these levels that lead to changes in the joint electron-nuclear state. A gate operation model defines the fidelity with which the prepared state compares to the intended results. Notably, the gate operation model is synthesized from the electrostatic fields and electron orbitals generated by the earlier stages of the computational workflow. This requires the model to account for the time varying magnetic field and the varying hyperfine coupling that drives the dynamics of the electron and nuclear spins. Its ultimate purpose is to provide insight and feedback into the control parameters that are necessary to demonstrate gate operation at a desired fidelity.
\par
\begin{figure}[t]
\centering
\includegraphics[width=2.5in]{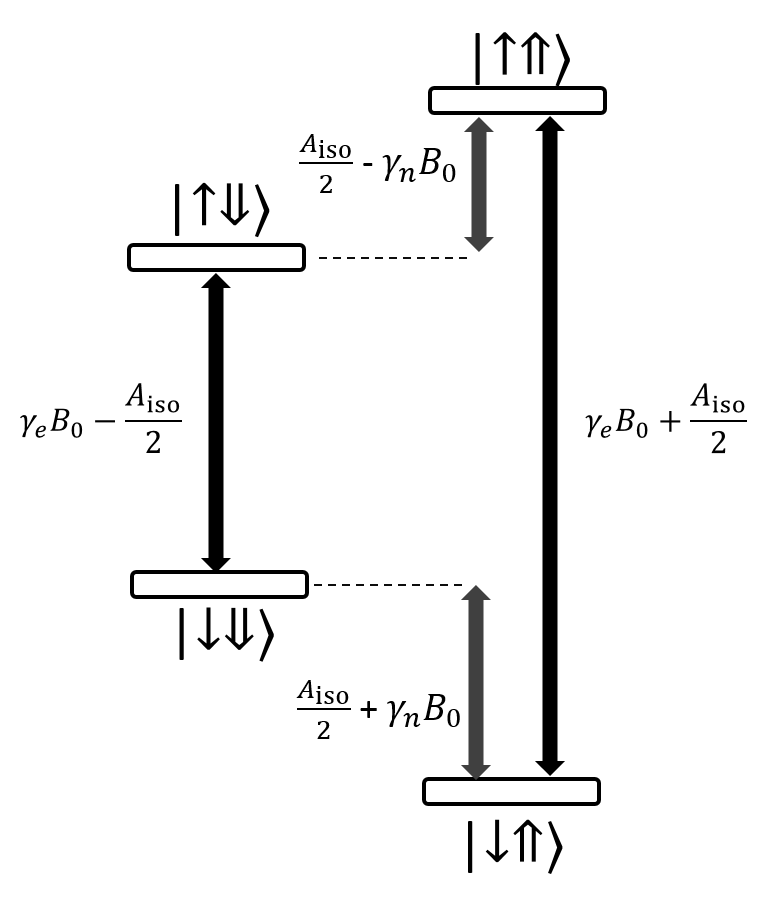}
\caption{Donor electron and nuclear spins states with their relevant energy separations not drawn to scale. Transitions between the electron $(|\uparrow\rangle, |\downarrow\rangle)$ or nuclear spin states $(|\Uparrow\rangle, |\Downarrow\rangle)$ are induced when $A_{\mathrm{iso}}$ is tuned such that the energy gap between the spin states matches the frequency $\omega$ of an externally applied oscillating magnetic field. We assume $\gamma_{n} B_{0} < A_{\mathrm{iso}}/2$.}
\label{fig:Spinlevels}
\end{figure}

A single-donor qubit gate operation model is defined by the interactions between the electron and nuclear spin states and the applied fields described by the time-dependent Hamiltonian
\begin{dmath}
H(t) = A_{\mathrm{iso}}(t) \mathbf{S} \cdot \mathbf{I} + \gamma_e B_{0} S_{z} - \gamma_n B_{0} I_{z}  + \gamma_e B_{ac} \cos(\omega t) S_{x} - \gamma_n B_{ac} \cos(\omega t) I_{x}
\label{eq:SpinHamiltonian}
\end{dmath}
where $\mathbf{S}$ is the electron spin operator and $S_{x,z}$ is its $x,z$-component,  $\mathbf{I}$ the nuclear spin operator and $I_{x,z}$ is its $x,z-$component, $B_{0}$ is the applied DC magnetic field, and $B_{ac}$ is the amplitude of the oscillating magnetic field used to drive the spin qubits at frequency $\omega$. The constants $\gamma_e =$ 28.025 GHz/T and $\gamma_n =$ 17.235 MHz/T  are the gyromagnetic ratios of the electron and nucleus respectively.  In the regime where $(\gamma_{e} + \gamma_{n}) B_0 \gg A$, the eigenstates for the combined electron-nuclear spin state are well represented by the four-level energy diagram shown in Fig.~\ref{fig:Spinlevels}. For example, in the device demonstrated previously by Laucht et al. \cite{Laucht2015}, the time-dependence of the Hamiltonian is reflected in (a) the time-varying hyperfine coupling $A(t)$, which is controlled through changes to the applied bias, and (b) the oscillating magnetic field used to drive the spins. Transitions between the nuclear and electron spin states can be induced when $A$ is tuned such that the resulting energy gap matches the frequency $\omega$ of the applied magnetic field $B_{ac}$. 
%%%%%%%%%%%%%%%%%%%%%%%%%%%%%%%%%%%%%%%%%%%%%%%%%%
%%%%%%%%%%%%%%%%%%%%%%%%%%%%%%%%%%%%%%%%%%%%%%%%%%
\par
The spin dynamics of the system can be obtained by solving the Schrodinger equation, or its noisy equivalent, the Master equation. For either the Schrodinger equation or its noisy equivalent, the state of the donor system is recovered by integrating over the time-dependent hyperfine coupling and magnetic field captured by Eq.~\ref{eq:SpinHamiltonian}. However, it is apparent from the gyromagnetic ratios that the time scale for electron dynamics is much shorter than the time scale for nuclear spin dynamics, i.e., approximately three orders of magnitude. Consequently,  it is essential to solve the equations of motion with sufficient temporal resolution. In particular, the much shorter time scale characterizing the electron spin must be adequately sampled to accurately simulate the complete joint spin state. This situation represents a simplified example of the multi-scale physics that arises in the simulation of heterogeneous coupled quantum systems. For our discussion of a single-qubit gate operation, the four-level system represented by Fig.~\ref{fig:Spinlevels} is relatively straightforward to solve using direct diagonalization of the time-dependent Hamiltonian. Consequently, the most significant computational complexity stems from integrating the underlying Schrodinger equation, cf. discussion above. Other methods can be applied to solving the differential equations defining changes to the qubit state. Finite element methods as well as finite differencing can be used to alleviate some of the problems with multi-scale resolution. There are a variety of numerical packages available for performing any of these calculations, including general commercial solvers like MATLAB and Mathematica as well as open source solutions like the QuTiP Python module.
\par
\begin{figure}
\includegraphics[width=\columnwidth]{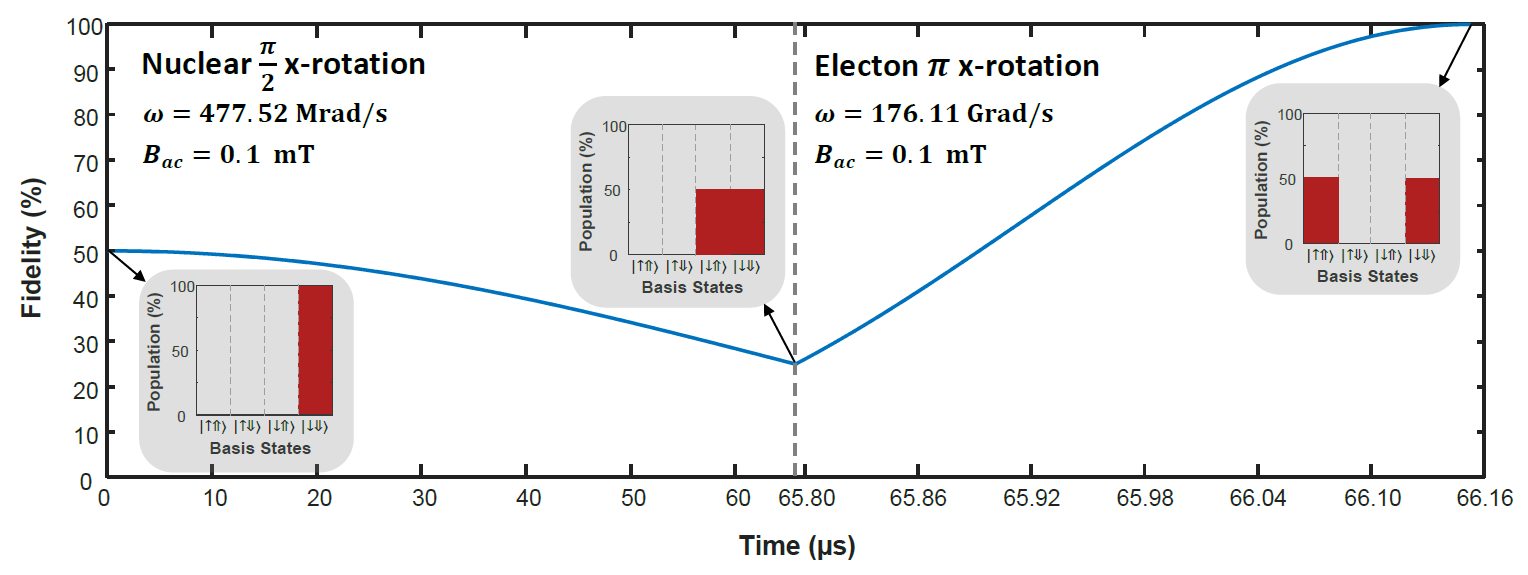}
\caption{Transient fidelity during simulated operation of creating an entangled Bell state between the electronic and nuclear spin states of a single donor atom. The system is initialized into the joint spin state $\ket{\downarrow,\Downarrow}$. A conditional $\pi/2$-rotation is applied to the nuclear spin state using NMR with frequency $\omega = A_{\mathrm{iso}}/2 + \gamma_n B_0 = 477.52$ Mrad/s. This is followed by a conditional $\pi$-rotation of the electron with ESR using $\omega = \gamma_e B_0+A_{\mathrm{iso}}/2$. The split x-axis changes resolution due to the distinct time scales required by the nuclear and electronic transitions. The simulated fidelities assume that the phase between the states are tracked and corrected accordingly. Prior to the rotation sequence described, $A_{\mathrm{iso}}$ must be tuned to 117.6 MHz by the gate electrodes in the nanostructure such that $\omega$ is on resonance with the transition frequencies shown in Fig.~\ref{fig:Spinlevels}.}
\label{fig:cnot}
\end{figure}

\revised{An example of how the gate operation model can be used to simulate the performance of a silicon donor qubits is shown in Fig.~\ref{fig:cnot}. This figure demonstrates preparation of a Bell entangled state using a sequence of NMR and ESR transitions for the joint electronic and nuclear spin state. In particular, we transform the joint spin state as
\begin{equation}
\ket{\downarrow,\Downarrow} \rightarrow \frac{1}{\sqrt{2}}\left(\ket{\downarrow,\Downarrow} + \ket{\uparrow,\Uparrow}\right)
\end{equation}
We realize this operation by applying a sequence of controlled rotations for the encoded qubits. We begin by assuming the electron-nuclear spin system is initialized as $\ket{\downarrow\Downarrow}$, which can be achieved using the techniques mentioned in Refs.~\cite{Pla2012,Pla2013}. We then apply a $\pi/2$-rotation to the nuclear spin state conditioned on the electron being in the state $\ket{\downarrow}$. This rotation is implemented using the $I_{x}$ interaction found in Eq.~(\ref{eq:SpinHamiltonian}) with $B_{ac}= 0.1$ mT and $\omega = A_{\mathrm{iso}}/2 + \gamma_n B_0$. The rotation takes approximately 65 $\mu$ s. The electron spin state is then conditionally rotated using the $S_{x}$ interaction and a second applied magnetic field $B_{ac}= 0.1$ mT  and $\omega = \gamma_e B_0 + A_{\mathrm{iso}}/2$. This rotation requires 360 ns to complete. Figure ~\ref{fig:cnot} illustrates that perfect fidelity is obtained when using these control conditions for an exactly characterized four-level system.}
%%%%%%%%%%%%%%%%%%%%%%%%%%%%%%%%%%%%%%%%%%%%%
%%%%%%%%%%%%%%%%%%%%%%%%%%%%%%%%%%%%%%%%%%%%%
%%%%%%%%%%%%%%%%%%%%%%%%%%%%%%%%%%%%%%%%%%%%%
%%%%%%%%%%%%%%%%%%%%%%%%%%%%%%%%%%%%%%%%%%%%%
\par
A more significant burden arises when modeling realistic devices that are marked by uncertainty in the device fabrication or operation. In particular, the statistical variability in a diverse range of parameters such as the donor position and other material imperfections as well as electromagnetic field noise and uncertainties in device fabrication must be taken into account. Whereas classical averaging of the quantum state representation can be performed by integrating over a known noise distribution, this may be approximated numerically by drawing samples for Monte Carlo analysis. However, the influence of these noise sources on the hyperfine coupling and, more generally, the electron orbital, are not directly evident. Instead, these noise models require averaging over the output of the electrostatic and materials models described in Secs.~\ref{sec:em} and \ref{sec:qm}. \revised{The need to repeat these calculations several times serves to greatly increase the computational complexity required for accurately simulating gate operations. As noted in Sec.~\ref{sec:qm}, the dependence of the hyperfine splitting (HFS) with location may be estimated based on expectations for the behavior of the electron wave function. Given a model for the distribution of donor location, the predicted insensitivity to lateral displacements and the near exponential decrease with depth can provide a more computationally efficient model for the distribution of possible HFS values. Accordingly, this analytic distribution for HFS can be used in place of repeated electronic structure calculations.
}
\par
\revised{Our model spin Hamiltonian in Eq.~(\ref{eq:SpinHamiltonian}) has the electron and nuclear spin states coupled via the hyperfine interaction $A_{\mathrm{iso}}(t)$. The hyperfine coupling is time dependent since it has to be tuned appropriately for gate operations, such that the frequency $\omega$ of the oscillating magnetic field matches the relevant energy separation between the different eigenstates in Fig.~\ref{fig:Spinlevels}. We highlight that the variation of $A_{\mathrm{iso}}(t)$ should be adiabatic, such that the excited donor orbital states are not populated during the process. Our computational chemistry methods described in the previous section do not permit for solving the time-dependent variation of $A_{\mathrm{iso}}(t)$ to quantify the adiabaticity. 
}
\par
\revised{
To maintain adiabaticty, the required timscale for varying $A_{\mathrm{iso}}(t)$ should be much longer than $h/E$, where $h$ is Planck's constant and $E$ is the energy separation between the donor ground and first excited states. In realistic nanostructures, $E$ takes a value of $\sim$ 10 meV  \cite{Mohiyaddin2013nl}. This implies that the hyperfine coupling should be varied at timescales much longer than $\sim$ 0.4 ps to ensure negligible population of the excited donor states. A recent experiment indeed showed that the hyperfine coupling could be tuned over several microseconds to demonstrate single-qubit operations \cite{Laucht2015}. While high-frequency noise in the gate voltages can also potentially cause non-adiabatic transitions, we emphasize that these experiments typically use attenuation filters (beyond MHz frequencies) on the gate electrodes that limit high frequency noise. Similarly, the noise that arises from intrinsic charge fluctuations in nanostructures typically have an inverse dependence on frequency \cite{Muhonen2014}, such that any contributions at terahertz frequencies or greater would be substantially smaller. As a result, we expect that it is reasonable to assume the electron orbital quickly equilibrates during the rise time of the applied voltage used to control $A_{\mathrm{iso}}(t)$. 
}
\par
\revised{
There is some uncertainty as to whether the dynamics arising from electron-phonon coupling and fluctuations arisign from thermal motion of the nuclei in the nanostructure may impact $A_{\mathrm{iso}}(t)$. While the low temperature $\sim$ 100 mK operation of such devices may rule out this possibility, it is possible to further investigate this situation by solving for the dynamics of electron orbital wave function using the time-dependent Schroedinger equation (TDSE). Generally, DFT and TB approaches allow for directly simulating excitation of the donor wave function in response to time-dependent fields/perturbation and thermal motion of nuclei. For silicon donors, the dynamics of the donor electron is especially significant since it serves as the handle by which logical operations are carried out, and the simulation of real-time electron dynamics using DFT is a promising tool. This requires solving the TDSE for electrons in the von Neumann density matrix representation as
\begin{equation}
\imath \hbar \frac{d\rho_{e}(t)}{dt} =  [H_{e}(t),\rho_{e}(t)]  
\label{eq:tdse}
\end{equation}
where $\rho_{e}(t)$ and $H_{e}(t)$ stand, respectively, for the time-dependent electronic density and electronic Hamiltonian, and the right hand side is the commutator $[H_{e},\rho_{e}]=H_{e}\rho_{e}-\rho_{e} H_{e}$.
The Hamiltonian $H_{e}(t)$ now describes mean field interactions of the donor electron with the nuclei, other electrons and external fluctuations. 
The interaction of the materials with external fields from the electrodes or other sources is included by adding time-dependent external potentials to the Hamiltonian. 
}
\par 
\revised{
The solution to Eq.~(\ref{eq:tdse}) is usually obtained from diagonalization of the Hamiltonian $H(t)$. This may be achieved by the formation of complex-valued exponential time-evolution operator,
\begin{equation}
U_{e}(t,t')=\exp\left[-\frac{\imath}{\hbar} \int_{t}^{t'}  H_{e}(t'')dt''\right],
\label{eq:U}
\end{equation}
from the eigenstates of the $H_{e}(t'')$-matrix, where $t'' \in [ t,t']$. 
The evolution operator $U(t,t')$ is then directly applied to the density matrix $\rho(t)$ and the new (propagated) density matrix at time time $t'$ is given by $\rho_{e}(t')  =  U_{e}(t,t')^\dag  \rho_{e}(t)  U_{e}(t,t')$, where $U(t,t')^\dag$ stands for a Hermitian conjugation of $U(t,t')$.  It is possible to implement the time-propagation using a first-order Magnus expansion and iterative diagonalization \cite{gpu,jingsong,jpcl,pccp}.  This procedure  for  the time dependent electron dynamics  in conjunction with tight-binding DFT has been proved to be very  efficient for molecular systems consisting thousands of atoms. The computational scaling is comparable with standard ground state calculations while allowing inclusion of nuclei motion and beyond linear response effect.
}
\par
\revised{
The set of coupled equations given by Eqs.~(\ref{eq:Aiso}), (\ref{eq:SpinHamiltonian}) and (\ref{eq:tdse}) capture the time-dependent dynamics of the electron and nuclear spin states. Whereas the time-dependence of the electron wave function can be modeled explicitly using Eq.~(\ref{eq:tdse}), the corresponding HFS can be calculated and substituted into the spin Hamiltonian given by Eq.~(\ref{eq:SpinHamiltonian}). The complete dynamics of the donor systems can be simulated. However, the large discrepancy in time scales for the electron and the spin dynamics relevant to logical computation make it clear that the granularity for these simulations must interpolate between these two extremes. Foremost, it is only necessary to simulate the electron dynamics on those timescales that induce changes. This includes the rise time for the applied voltage and the correlation time for the electronic noise. Outside of these timescales, we may approximate the HFS value by a constant.
}
\par
Ultimately, our goal of modeling gate operation in realistic silicon donor devices has emphasized the coupled nature of electromagnetic and material modeling. These simulations produce numerical representations of the instantaneous quantum state that can be used to compare against the expected state dynamics and experimental observables. The resulting fidelity, as measured by the overlap between the observed and expected state values, is a useful quantity for assessing the quality of a given design, material sample, or control sequence. In addition, the gate operation model can eventually be validated against experimental samples by making comparison between measured fidelities and those calculated under the model. However, the importance of high-fidelity models and accurately dynamical simulation must be stressed, since these models can then provide feedback for the improved design of new devices.
%%%%%%%%%%%%%%%%%%%%%%%%%%%%%%%%%%%%%%%%%%
\section{Conclusions}
\label{sec:con}
Given recent advances in the demonstration of prototype quantum computing device, new tools are needed for modeling and simulation of the fabricated qubits. Silicon donor qubits offer an example of how a multi-staged modeling workflow is needed to synthesize electrical and quantum physical properties into a single consistent model. Our multi-physics approach integrates models of the electrostatics, quantum chemistry, and device operation into a common framework. We must caution that the scaling of direct simulation is very unfavorable for quantum mechanical systems. In particular, we do not expect DFT calculations to remain tractable for more than two donor atoms separated by more than a few nanometers. However, alternative approaches like TB and DFTB will be useful for simulating larger systems. We believe that the modeling and simulation workflow presented here for silicon donor devices will provide valuable insights into the ongoing development of quantum computing devices that require atomistic control. We expect the silicon donor TCAD tool to be essential for maturing silicon donor systems into functioning qubits, including integration into the device fabrication process. 
%%%%%%%%%%%%%%%%%%%%%%%%%%%%%%%%%%%%%%%%%%
\section*{Acknowledgments}
The authors thank Arne Laucht of University of New South Wales for help in specifying the device design in Fig.~\ref{fig:comsol_device}. A portion of this research was conducted at the Center for Nanophase Materials Sciences, which is a DOE Office of Science User Facility.  NCN/nanohub.org computational resources funded by the National Science Foundation under contract number EEC-1227110 were for the NEMO-3D simulations. The XSEDE allocation TG-DMR110037 is also acknowledged.​ This manuscript has been authored by UT-Battelle, LLC, under Contract No. DE-AC0500OR22725 with the U.S. Department of Energy. The United States Government retains and the publisher, by accepting the article for publication, acknowledges that the United States Government retains a non-exclusive, paid-up, irrevocable, world-wide license to publish or reproduce the published form of this manuscript, or allow others to do so, for the United States Government purposes. The Department of Energy will provide public access to these results of federally sponsored research in accordance with the DOE Public Access Plan.
%%%%%%%%%%%%%%%%%%%%%%%%%%%%%%%%%%%%%%%%%%%%%%%%%%
\section*{References}
\bibliography{nano2016}
\bibliographystyle{iopart-num}

\end{document}